\documentclass[%
 reprint,
superscriptaddress,
 amsmath,amssymb,
 aps,
 onecolumn,
]{revtex4-1}

\usepackage{graphicx}
\usepackage{dcolumn}
\usepackage{bm}

\begin{document}

\preprint{APS/123-QED}

\title{Implementation of a Measurement-Device-Independent Entanglement Witness}

\author{Ping Xu}
\affiliation{Shanghai Branch, National Laboratory for Physical Sciences at Microscale and Department of Modern Physics, University of Science and Technology of China, Shanghai, China}
\affiliation{Synergetic Innovation Center of Quantum Information \& Quantum Physics,  University of Science and Technology of China, Hefei, Anhui, China}

\author{Xiao Yuan}
\affiliation{Center for Quantum Information, Institute for Interdisciplinary Information Sciences, Tsinghua University, Beijing, China}

\author{Luo-Kan Chen}

\author{He Lu}

\author{Xing-Can Yao}
\affiliation{Shanghai Branch, National Laboratory for Physical Sciences at Microscale and Department of Modern Physics, University of Science and Technology of China, Shanghai, China}
\affiliation{Synergetic Innovation Center of Quantum Information \& Quantum Physics,  University of Science and Technology of China, Hefei, Anhui, China}

\author{Xiongfeng Ma}
\affiliation{Center for Quantum Information, Institute for Interdisciplinary Information Sciences, Tsinghua University, Beijing, China}

\author{Yu-Ao Chen}

\author{Jian-Wei Pan}
\affiliation{Shanghai Branch, National Laboratory for Physical Sciences at Microscale and Department of Modern Physics, University of Science and Technology of China, Shanghai, China}
\affiliation{Synergetic Innovation Center of Quantum Information \& Quantum Physics,  University of Science and Technology of China, Hefei, Anhui, China}


\date{\today}

\begin{abstract}
Entanglement, the essential resource in quantum information processing, should be witnessed in many tasks such as quantum computing and quantum communication. The conventional entanglement witness method, relying on an idealized implementation of measurements, could wrongly conclude a separable state to be entangled due to imperfect detections. Inspired by the idea of a time-shift attack, we construct an attack on the conventional entanglement witness process and demonstrate that a separable state can be falsely identified to be entangled. To close such detection loopholes, based on a recently proposed measurement-device-independent entanglement witness method, we design and experimentally demonstrate a measurement-device-independent entanglement witness for a variety of two-qubit states. By the new scheme, we show that an entanglement witness can be realized without detection loopholes.
\end{abstract}

\maketitle


Quantum entanglement plays an important role in the nonclassical phenomenons of quantum mechanics. Being the key resource for many tasks in quantum information processing, such as quantum computation \cite{wiesner}, quantum teleportation \cite{tele}, and quantum cryptography \cite{bb84,Ekert91}, entanglement needs to be verified in many scenarios. There are several proposals to witness entanglement and we refer to Ref. \cite{guhne2009} for a detailed review. A conventional way to detect entanglement, the entanglement witness (EW), gives one of two outcomes: ``Yes'' or ``No'', corresponding to the conclusive result that the state is entangled or to failure to draw a conclusion, respectively.  Mathematically, for a given entangled quantum state $\rho$, a Hermitian operator $W$ is called a witness, if $tr[W\rho] < 0$ (output of `Yes') and $tr[W\sigma]\geq0$ (output of `No') for any separable state $\sigma$. Note that there could also exist an entangled state $\rho'$ such that $tr[W\rho']\geq0$ (output of `No'). In the experimental verification, one can realize the conventional EW with only local measurements by decomposing $W$ into a linear combination of product Hermitian observables \cite{guhne2009}.

\begin{figure}[bht]
\centering
\resizebox{8cm}{!}{\includegraphics[scale=1]{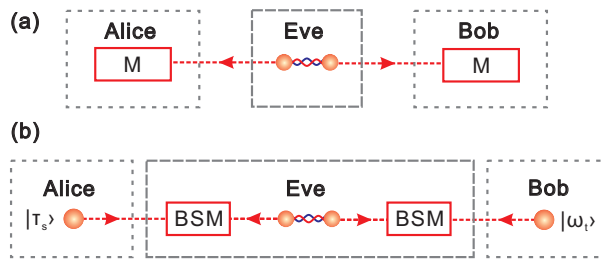}}
\caption{(a) Conventional EW setup, where Alice and Bob perform local measurements separately and collect information to decide whether the input state is entangled or not. (b) Measurement-device-independent (MDI) EW setup, where Alice and Bob each prepare an ancillary state and a third party Eve performs Bell state measurements (BSMs) on the ancillary states and the to-be-witnessed bipartite state. Based on the choices of Alice and Bob's ancillary states and the BSM results, they can judge whether the input state is entangled or not.}\label{fig:EWAMDIEW}
\end{figure}

Focusing on the bipartite scenario, a general illustration of the conventional EW is shown in Fig.~\ref{fig:EWAMDIEW}(a), where two parties, Alice and Bob, each receive one component of a bipartite state $\rho_{AB}$ from an untrusted third party Eve. They want to verify whether $\rho_{AB}$ is entangled or not, by performing local operations and measurements on $\rho_{A}=Tr_B[\rho_{AB}]$ and $\rho_{B}=Tr_A[\rho_{AB}]$. The correctness of such witness relies on implementation details of $W$. An unfaithful implementation of $W$, say, due to device imperfections, would render the witness results unreliable. For example, the measurement devices used by Alice and Bob might possibly be manufactured by another untrusted party, who could collaborate with Eve and deliberately fabricate devices to make the real implementation $W' = W+\delta W$  deviate from $W$, such that $W'$ is not a witness any more,
\begin{equation}\label{MDIEW:Def:SuccessAttack}
  tr[W'\sigma]<0<tr[W\sigma].
\end{equation}
That is, with the deviated witness $W'$, a separable state $\sigma$ could be identified as an entangled one, which is more likely to happen when $tr[W\sigma]$ is near zero.

There is a strong similarity between the EW and the quantum key distribution (QKD) where an entanglement-breaking channel would cause insecurity \cite{Lutkenhaus04}. Roughly speaking, it is crucial for Alice and Bob to prove that entanglement can be preserved in a secure QKD channel. From this point of view, there exists a correlation between the security of the QKD and the success of the EW. For the varieties of attacks in the QKD, such as time-shift attacks \cite{qi07} and fake-state attacks \cite{Makarov06}, one may also find similar detection loopholes in the conventional EW process. Originating from this analogy, we construct a time-shift attack that manipulates the efficiency mismatch between detectors used in an EW process. Under this attack, any state could be witnessed to be entangled, even if the input state is separable. By this example, we demonstrate that there do exist loopholes in the conventional EW procedure.

Recently, Lo et al. \cite{Lo12} proposed a measurement-device-independent (MDI) QKD method, which is immune to all hacking strategies on detection. Due to the similarity between the QKD and the EW, one would also expect that there exist EW schemes without detection loopholes. Meanwhile, a nonlocal game is proposed to distinguish any entangled state from all separable states \cite{Buscemi12}. Inspired by this game, Branciard et al. \cite{Branciard13} proposed an MDIEW method, where they proved that there always exists an MDIEW for any entangled state with untrusted measurement apparatuses.

As shown in Fig.~\ref{fig:EWAMDIEW}(b), Alice and Bob want to identify whether a given bipartite state, prepared by an untrusted party Eve, is entangled or not without trusting measurement devices. To do so, Alice (Bob) prepares an ancillary state $\tau_s$ ($\omega_t$) and sends it along with the to-be-witnessed bipartite state to a willing participant, who can be assumed to be Eve again in the worst case scenario. Eve performs two Bell-state measurements (BSMs) on the two ancillary states and the bipartite state. Then, she announces to Alice and Bob the results of BSMs, based on which they will witness the entanglement of the bipartite state. In the MDIEW, it is guaranteed that a separable state will never be wrongly identified as an entangled one, even if Eve maliciously makes wrong measurements and/or announces unfaithful information \cite{Branciard13}.



In the experiment, we first show an example of the time-shift attack on the conventional EW process and demonstrate how a separable state can be falsely identified to be entangled when a large efficiency mismatch happens. Then we design and experimentally realize an MDIEW scheme to close such detection loopholes. The MDIEW is used to testify the entanglement of various bipartite states starting from maximally entangled to separable ones.  Note that we use heralded single-photon sources to prepare the two ancillary states; thus, our demonstration is realized by a six-photon interferometry.

\begin{figure}[bht]
\centering
\resizebox{8cm}{!}{\includegraphics[scale=1]{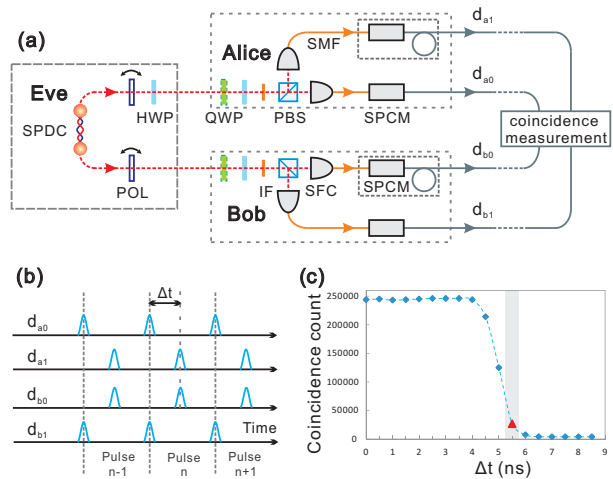}}
\caption{Time shift attack on the  conventional EW. (a) Experimental setup of the time-shift attack. Photon pairs are generated by SPDC using a femtosecond pump laser with a central wavelength of 390 nm and a repetition frequency of 80 MHz. POL: polarizer, HWP: half-wave plate, QWP: quarter-wave plate, IF: interference filter with 780 nm central wavelength, PBS: polarizing beam splitter, SFC: single-mode fiber coupler, SMF: single-mode fiber, SPCM: single-photon-counting module, some with extra internal delay lines.
(b) Synchronization between SPCMs.  Build-in delay lines enable Eve to shift the output signals $d_{a1}$ and $d_{b0}$ by $\Delta t$.
(c) Coincidence count versus time delay, where the time window is set to 4 ns. All data points are measured for 2 seconds, and the time-shift attack is implemented with $\Delta t=5.50 \pm0.24$ ns, which corresponds to the grey area.}\label{TSAS}
\end{figure}

\textbf{Time-shift attack}, originated from quantum cryptography \cite{qi07}, takes advantage of the efficiency mismatch of the measurement devices. As shown in Fig.~\ref{TSAS}(a), typically two detectors are used on each side of Alice and Bob. By controlling the single-photon-counting modules (SPCMs) and coincidence gate, Eve is able to enlarge the efficiency mismatch and hence manipulate the EW result.

To implement this attack, we choose a conventional witness
\begin{equation*}\label{eq:w}
  W = \frac{1}{2}I-|\Psi^-\rangle\langle\Psi^-|,
\end{equation*}
for bipartite states in the form of
\begin{equation}\label{eq:rho}
\rho^v_{AB}=(1-v)|\Psi^-\rangle\langle\Psi^-|+\frac{v}{2}(|HH\rangle\langle HH| + |VV\rangle\langle VV|),
\end{equation}
where $H$ ($V$) denotes the horizontal (vertical) polarization of the single photons and $|\Psi^{-}\rangle = (|HV\rangle - |VH\rangle)/\sqrt{2}$ is a Bell state. By decomposing $W$ into a linear combination of product Pauli matrices, the EW can be realized by local measurements,
\begin{equation*}\label{eq:expEW}
Tr[W\rho_{AB}]=\frac{1}{4}\left( {1 + \left\langle {{\sigma _x}{\sigma _x}} \right\rangle  + \left\langle {{\sigma _y}{\sigma _y}} \right\rangle  + \left\langle {{\sigma _z}{\sigma _z}} \right\rangle } \right).
\end{equation*}
That is, to identify the entanglement, Alice and Bob just have to each analyze the qubit state in three bases separately. When the bipartite state is projected to the positive (negative) eigenstates of $\sigma_x\sigma_x$, $\sigma_y\sigma_y$, and $\sigma_z\sigma_z$, it will contribute positively (negatively) to the witness result $Tr[W\rho_{AB}]$. For example, when measuring ${\sigma _x}{\sigma_x}$, Alice and Bob will both project the input state to the eigenstates of $\sigma_x$, $\sigma_x^+$ or $\sigma_x^-$, with corresponding eigenvalues of $+1$ or $-1$, respectively, and obtain probabilities $\left\langle {{\sigma _x^\pm}{\sigma _x^\pm}} \right\rangle$. Then the value of $\left\langle {{\sigma _x}{\sigma _x}} \right\rangle$ is defined as $\left\langle {{\sigma _x^+}{\sigma _x^+}} \right\rangle + \left\langle {{\sigma _x^-}{\sigma _x^-}} \right\rangle - \left\langle {{\sigma _x^+}{\sigma _x^-}} \right\rangle - \left\langle {{\sigma _x^-}{\sigma _x^+}} \right\rangle$.  From Eve's point of view, she wants to convince Alice and Bob that the bipartite state is entangled, that is, $Tr[W\rho_{AB}]<0$. Thus, her objective is to suppress the positive contributions of $Tr[W\rho_{AB}]$, such as $\left\langle {{\sigma _x^+}{\sigma _x^+}} \right\rangle$ and $\left\langle {{\sigma _x^-}{\sigma _x^-}} \right\rangle$ for the ${\sigma _x}{\sigma_x}$ measurement, by manipulating the coincidence rate between SPCMs, equivalently enlarging the detector efficiency mismatch. In this case, from Alice and Bob's point of view, the real implemented witness $W'$ is deviated from the desired one $W$, and satisfies Eq.~\eqref{MDIEW:Def:SuccessAttack}. More details of the time-shift attack can be found in Appendix.


In our experiment, as shown in Fig.~\ref{TSAS}(a), by encoding qubits in the polarization of photons, the bipartite state $(\vert HH\rangle_{ab}+\vert VV\rangle_{ab})/\sqrt 2$ is generated via spontaneous parametric down conversion (SPDC). Two adjustable POLs are used to disentangle the initial state and project it to $\vert HH \rangle_{ab}$ and $\vert VV \rangle_{ab}$ with equal probabilities, corresponding to the separable state with $v = 1$ in Eq.~\eqref{eq:rho}. After a $45^\circ$ HWP, the to-be-witnessed two-qubit system is prepared in the state of $\rho_{AB} = \left( {{{\left| {HV} \right\rangle }}\left\langle {HV} \right| + {{\left| {VH} \right\rangle }}\left\langle {VH} \right|} \right)/2$. Then Alice and Bob each perform polarization analysis on a qubit from the bipartite state using waveplates, PBSs and SPCMs, and guide the electronic signals from the SPCMs into a coincidence gate.

As shown in Fig.~\ref{TSAS}(b), in the time-shift attack, Eve controls the delay lines in the detection systems and the time window of the coincidence gate, and hence, manipulates the time-dependent coincidence counting rates between detectors $d_{a0}$ and $d_{b0}$, $d_{a1}$ and $d_{b1}$. Hence, she can suppress the positive contributions of measurements $\left\langle {{\sigma _x}{\sigma _x}} \right\rangle , \left\langle {{\sigma _y}{\sigma _y}} \right\rangle $ and $\left\langle {{\sigma _z}{\sigma _z}} \right\rangle $. In our demonstration, by setting proper parameters, we let the positive contributions drop to 10.9(1) $\%$ of their original values.
Since this attack would not affect the negative contributions of $Tr[W\rho_{AB}]$, the experimental outcomes for $\left\langle {{\sigma _x}{\sigma _x}} \right\rangle , \left\langle {{\sigma _y}{\sigma _y}} \right\rangle $ and $\left\langle {{\sigma _z}{\sigma _z}} \right\rangle $ become negative as expected.  Finally, Alice and Bob obtain a witness of $\rho_{AB} $ be $ tr\left[ {W'\rho_{AB} } \right] =  - 0.379\left( 4 \right)$,  although the input state $\rho_{AB} $  is, in fact, separable. By changing $\Delta t$ to a larger value, one can even obtain a fake result for that from a maximal entangled state. Thus, a separable bipartite state could be wrongly witnessed to be entangled when Eve is able to manipulate the detection system. It is not hard to see that for any state $\rho$, Eve can perform a similar attack and trick Alice and Bob into thinking that it is entangled.


Note that in the original time-shift attack in the QKD \cite{qi07}, Eve is only able to partially control the detection efficiency by manipulating the timing of the quantum signals. In that case, Eve cannot arbitrarily enlarge the efficiency mismatch between desired and undesired detection events. In the EW case, there are two quantum signals Eve can manipulate. From our demonstration, we show that by controlling the coincident gates, Eve is able to arbitrarily decrease the coincident detection efficiency (down to 0) for any type of detection events. Thus, Eve can make the EW device output any of her desired results. From this point of view, the efficiency mismatch problem is more serious in the EW.

\textbf{MDIEW} is able to close all loopholes introduced by imperfect measurement devices. In this scheme, to witness entanglement existing in a bipartite state $\rho_{AB}$, Alice and Bob randomly choose and prepare ancillary states $\tau_s$ and $\omega_t$ from state sets $\{\tau_s\},\{\omega_t\}$, respectively. By performing two BSMs on the ancillary states and the bipartite state $\rho_{AB}$ as shown in Fig.~\ref{fig:EWAMDIEW}(b), conditional probabilities $p(a,b|\tau_s, \omega_t) = \textrm{Tr}[(M^a\otimes M^b)(\tau_s\otimes\rho_{AB}\otimes\omega_t)]$ are obtained, where $M^a(M^b)$ denotes the positive operator-valued measure (POVM) element of Eve's BSM with outcome $a(b)$. The convex combination of the probabilities $p(a,b|\tau_s, \omega_t)$
\begin{equation}\label{MDIdef}
 J(\rho_{AB}) = \sum_{a,b,s,t}\beta^{a,b}_{s,t}p(a, b|\tau_s, \omega_t)
\end{equation}
define an MDIEW. That is, $\rho_{AB}$ is entangled while $J(\rho_{AB})<0$ and for any separable state $\sigma_{AB}$, we have $J(\sigma_{AB})\geq0$.

For any entangled state $\rho_{AB}$ and its conventional witness $W$, one can construct a MDIEW in the form of Eq.~\eqref{MDIdef} by decomposing $W$ as a linear combination of product Hermitian operators, $\{\tau_s\otimes\omega_t$\}, which are used as the density matrices of the ancillary states \cite{Branciard13}. The coefficients $\beta$ depend on $W$, the outcomes of the BSMs, and ancillary states. We leave the calculation of $\beta$ to Appendix.

\begin{figure}[hbt]
\centering
\resizebox{8cm}{!}{\includegraphics[scale=1]{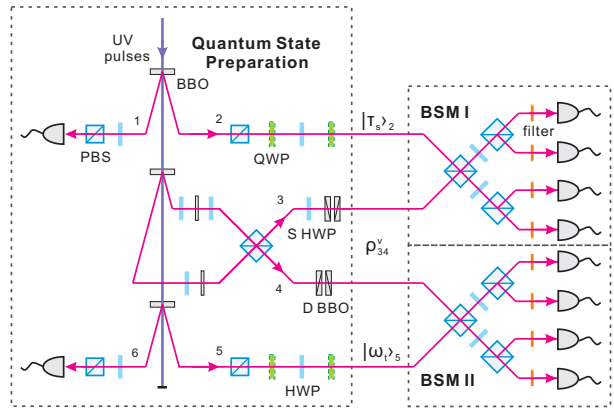}}
\caption{Experimental setup for the MDIEW. The photon pairs are generated by type-II SPDC in 2-mm $\beta$-barium-borate (BBO) crystals. The pulsed pump laser has a central wavelength of 390 nm and a repetition rate of 76 MHz. To prepare the desired state~\eqref{eq:rho}, two 2-mm decoherer BBOs (D BBO) on each side with fast axis setting at $0^\circ $ (up) and $180^\circ$ (down) to reduce the spatial walk-off effect. By changing the angle $\theta$ of the selector HWP (S HWP), the desired state~\eqref{eq:rho} is prepared with $ v=cos^2(2 \theta)$. Heralded photons 2 and 5 are triggered by the detections of photon 1 and 6, respectively. Waveplates are used to rotate the polarizations to encode photons 2 and 5 to the desired states, $\left| {{\tau _s}}\right\rangle _2$ and $\left| {{\omega _t}} \right\rangle _5$. The BSM module is composed of three PBSs and two HWPs at $22.5^\circ $. All photons are filtered by narrow-band filters (with $\lambda_{FWHM}$  = 2.8 nm for BSM I and $\lambda _{FWHM}$  = 8.0 nm for BSM II) and then coupled into single-mode fibers which connect to SPCMs.
}\label{expsetup}
\end{figure}

Our experimental setup for MDIEW is shown in Fig.~\ref{expsetup}, where a six-photon interferometry is utilized. The to-be-witnessed bipartite state $\rho^v_{34}$, defined in Eq.~\eqref{eq:rho}, is encoded in the photon pair 3 and 4. Photon pairs 1, 2 and 5, 6 are used to prepare the ancillary input states $|\tau_s\rangle_2$ and $|\omega_t\rangle_5$, respectively. In our work, various bipartite states $\{\rho^v_{34}\}$, from maximally entangled to separable, are prepared and tested with the MDIEW. The bipartite state $\rho^v_{34}$ is first prepared in the Bell state $\left| {{\Phi ^ - }} \right\rangle _{34} = \left( {{\left| {HH} \right\rangle } - \left| {VV} \right\rangle } \right)/\sqrt 2$ via a Bell-state synthesizer \cite{yao12}. As the coherence length of photons is limited by the interference filtering, two 2-mm BBO crystals in each arm result in a relative phase delay between horizontal and vertical polarization components and cause polarization decoherence. Different $v$ can be selected by the ``state selector" \cite{White01}. They satisfy the relation
\begin{equation}\label{eq:vtheta}
v  = {\cos ^2}\left( {2\theta } \right),
\end{equation}
where $\theta$ is the angle of the fast axis of the selector HWP.

In the experiment, eight ancillary state pairs $\{\tau_s, \omega_t\}$ are prepared. The states are encoded by tunable waveplates (one HWP sandwiched by two QWPs), which can realize arbitrary single-qubit unitary transformation. Different from direct polarization measurement in the conventional EW, the analysis of MDIEW is completed by BSMs on $\rho_{3}^v\otimes|\tau_s\rangle\langle\tau_s|_2$ and $\rho_{4}^v\otimes|\omega_t\rangle\langle\omega_t|_5$, with two, $|\Phi^\pm\rangle=(|HH\rangle \pm |VV\rangle)/{\sqrt{2}}$,  out of four outcomes being collected.

\begin{figure}[hbt]
\centering
\resizebox{8cm}{!}{\includegraphics[scale=0.6]{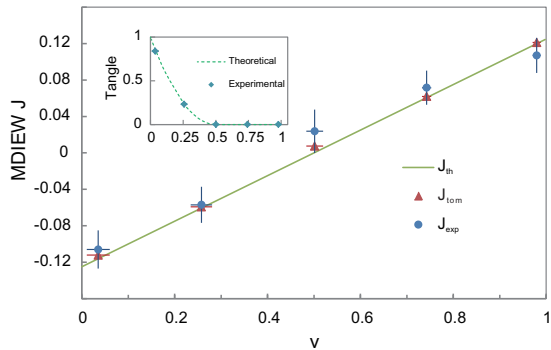}}
\caption{MDIEW values are compared for three cases. The theoretical results ($J_{th}$, solid line) are calculated for the states $\rho_{AB}^v$ with different values of $v$ in Eq.~\eqref{eq:rho}. The tomography results ($J_{tom}$, triangle points) are evaluated for the states $\rho_{34}^v$ after performing tomography on the to-be-witnessed bipartite state. Each point of the experimental results ($J_{exp}$, circular points) is measured from a 16-hour experiment. Vertical error bars indicate one standard deviation and horizontal error bars of the fitting values $v$ from state tomography are described in Appendix. The inset shows theoretical and experimental values of tangle for input states $\rho_{34}^v$.} \label{resmdi}
\end{figure}

As defined in Eq.~\eqref{MDIdef}, we obtain the experimental results $J_{exp}^v$ as shown in Fig.~\ref{resmdi}. In comparison, we also plot $J_{th}(\rho_{AB}^v)$ for all values of $v$. Recall that in the aforementioned time-shift attack demonstration, the conclusion from the conventional witness is entangled for $v=1$, whereas here we show that our MDIEW result is 0.107 $\pm$ 0.019 and does not conclude an entangled state. One can see that our MDIEW is immune to this attack. The BSM results only provide as information whether or not the entanglement is successfully swapped. It is the ancillary states that determine whether the detection event contributes positively or negatively to the witness value defined in Eq.~\eqref{MDIdef}. Thus, by knowing and/or manipulating the BSM results, Eve cannot suppress the positive components of the witness, nor can she render the MDIEW to false conclusions.

Furthermore, we perform tomography on the to-be-witnessed bipartite states $\{\rho_{34}^{v}\}$. The results of the density matrices are shown in Appendix.
The values of $v$ are also fitted according to Eq.~\eqref{eq:vtheta} in Appendix, which are consistent with tomography results. We evaluate the MDIEW results, Eq.~\eqref{MDIdef}, from the results of the state tomography $J_{tom}$ as shown in Fig.~\ref{resmdi}. Meanwhile, to quantify the entanglement of the bipartite states $\{\rho_{34}^{v}\}$, we adopt the measure of tangle \cite{Wootters}, which can be directly calculated from tomography results. When the tangle goes to zero, the bipartite state becomes a separable state. As shown in the insert of Fig.~\ref{resmdi}, no entanglement exists when $v$ grows beyond $1/2$. Such a phenomenon is related to the ``sudden death of entanglement" \cite{PhysRevLett.93.140404}.




In summary, we show that the conventional EW is unconfident due to the loopholes on detections. Meanwhile, as a countermeasure, we design and implement the MDIEW for the bipartite scenario, which is immune to all detection loopholes. The experimental results show that the MDIEW is practical for real-life implementation. Our method can be extended to other multipartite quantum tasks, such as quantum secret sharing.


\begin{acknowledgments}
We acknowledge insightful discussions with Y.-J.~Deng and Z.~Zhang. This work has been supported by the National Basic Research Program of China Grants No.~2011CB921300, No.~2013CB336800, No.~2011CBA00300, and No.~2011CBA00301, the National Natural Science Foundation of China Grants, and the Chinese Academy of Sciences. P.~X.~and X.~Y.~contributed equally to this work.
\end{acknowledgments}

\begin{appendix}

\section{MDIEW}\label{MDIEWCAL}
Measurement-device-independent entanglement witness (MDIEW) provides means to witness entanglement of a quantum state without trusting measurement devices \cite{Branciard13}. The idea of MDIEW is inspired from the MDI quantum key distribution (MDIQKD) \cite{Lo12}. As proved in Ref.~\cite{Branciard13}, there always exists an MDIEW for any quantum state $\rho$, as one can always construct MDIEW based on the conventional witness $W$ which exists for any quantum state (we refer to \cite{guhne2009} for details of conventional entanglement witness). In the following, we will design an MDIEW scheme and apply it to a type of bipartite quantum states in the form of
\begin{equation}\label{rhovab}
   \rho^v_{AB}=(1-v)|\Psi^-\rangle\langle\Psi^-|+\frac v2(|00\rangle\langle00| + |11\rangle\langle11|),
\end{equation}
with $v\in[0,1]$ and $|\Psi^-\rangle=(|01\rangle-|10\rangle)/\sqrt{2}$. The state is entangled if $v<1/2$, which can be witnessed by a conventional EW,
\begin{equation}\label{wit}
W=\frac{1}{2}I-|\Psi^-\rangle\langle\Psi^-|,
\end{equation}
and its result, $tr[W\rho^v_{AB}] = (2v-1)/2$.

Practically, the conventional EW can be realized with only local measurements by decomposing $W$ into a linear combination of product Hermitian observables. In the bipartite scenario of Alice and Bob, they only need to perform local measurements to decide the entanglement of quantum states. In contrast, MDIEW requires Alice (Bob) to prepare another ancillary state ${\tau_s}$ (${\omega_t}$) and perform Bell-state measurements (BSMs) on the to be witnessed state and the ancillary state. Conditioned on the measurement outcomes, $a$ and $b$, MDIEW is defined as
\begin{equation}\label{Supp:MDIEWJ}
 J(\rho_{AB}) = \sum_{s,t}\beta^{a,b}_{s,t}p(a, b|\tau_s, \omega_t),
\end{equation}
where the choice of the ancillary states are labeled by $s$ and $t$. That is, $\rho_{AB}$ is entangled while $J(\rho_{AB})<0$ and for any separable state $\sigma_{AB}$, we have $J(\sigma_{AB})\geq0$.
Here the probabilities $p(a, b|\tau_s, \omega_t)$ are obtained from performing two BSMs on the to be witnessed state $\rho_{AB}$ and the ancillary states ${\tau_s}$ and ${\omega_t}$. That is,
\begin{equation}\label{eq:P}
  p(a, b|\tau_s, \omega_t) = Tr[(M_a\otimes M_b)(\tau_s\otimes\rho_{AB}\otimes\omega_s)],
\end{equation}
where $M_a$ and $M_b$ represent BSMs performed by Alice and Bob with outcome $a$ and $b$, respectively. In Eq.~\eqref{Supp:MDIEWJ}, the coefficient $\beta^{a,b}_{s,t}$ is determined by the choice of ancillary states, measurement outcomes and the conventional witness $W$. In the experiment, as only two $|\Phi^+\rangle=(|00\rangle + |11\rangle)/{\sqrt{2}}$ and $|\Phi^-\rangle=(|00\rangle - |11\rangle)/{\sqrt{2}}$ out of four BSM outcomes are recorded, we consider the outcomes of $a$ and $b$ to be $+$ and $-$, which refer to  $|\Phi^+\rangle$ and $|\Phi^-\rangle$, respectively. There are four kinds of $\beta^{a,b}_{s,t}$, depending on different values of $a$ and $b$. In the following, we will design $\beta^{a,b}_{s,t}$ for our MDIEW.

The case of $a =+$ and $b = +$ is considered in Ref.~\cite{Branciard13}. Decompose a conventional EW as a linear combination of product Hermitian operators, $\{\tau_s\otimes\omega_t$\},
\begin{equation}\label{decom1}
  W^{}=\sum_{s,t}\beta^{++}_{s,t}\tau_s^T\otimes\omega_t^T,
\end{equation}
where the superscript $T$ means matrix transpose. In the corresponding MDIEW, Alice and Bob prepare their ancillary states into $\{\tau_s\}$ and $\{\omega_t\}$, respectively. According to Eq.~\eqref{eq:P}, $p(+, +|\tau_s, \omega_t)$ is obtained by projecting the joint states $tr_B[\rho_{AB}]\otimes\tau_s$ and $tr_A[\rho_{AB}]\otimes\omega_t$ to the maximally entangled states $|\Phi^+_{AA}\rangle=(|00\rangle + |11\rangle)/{\sqrt{2}}$ and $|\Phi^+_{BB}\rangle=(|00\rangle + |11\rangle)/\sqrt{2}$, respectively. Then it is easy to show that the relation between MDIEW and the conventional EW is
\begin{equation}\label{}
  J(\rho_{AB}) = tr[W^{}\rho_{AB}]/4,
\end{equation}
which equals $({2v-1})/{8}$ using Eq.~\eqref{rhovab} and \eqref{wit}.

In our work, we also consider other BSM outcomes. For example, if Alice and Bob get outcomes $a = -$ and $b = -$, then $\beta^{--}_{s,t}$ is calculated similarly as Eq.~\eqref{decom1} by decomposing $W$,
\begin{equation}\label{}
W^{}=\sum_{s,t}\beta^{--}_{s,t}\tilde{\tau}_s^T\otimes\tilde{\omega}_t^T,
\end{equation}
where $\langle j|\tilde{\tau}|i\rangle = (-)^{i+j}\langle j|{\tau}|i\rangle$ and $\langle j|\tilde{\omega}|i\rangle = (-)^{i+j}\langle j|{\omega}|i\rangle$. By redefining the basis that $W$ is decomposed, $\{\tilde{\tau}\otimes\tilde{\omega}\}$, the ancillary states prepared by Alice and Bob are still $\{\tau_s\}$ and $\{\omega_t\}$. In this case, $p(-, -|\tau_s, \omega_t)$ is obtained by projecting the joint states $tr_B[\rho_{AB}]\otimes\tau_s$ and $tr_A[\rho_{AB}]\otimes\omega_t$ to the maximally entangled states $|\Phi^-_{AA}\rangle=(|00\rangle - |11\rangle)/{\sqrt{2}}$ and $|\Phi^-_{BB}\rangle=(|00\rangle - |11\rangle)/\sqrt{2}$, respectively.

With a similar manner, one can also decompose $W$ for the cases of $a =+$ and $b = -$, $a=-$ and $b = +$. All the four cases of $a$ and $b$ are summarized in Table \ref{Tab:Wexpansion}.

\begin{table}[hbt]
\caption{Decomposition of $W$ based on different measurement outcomes.} \label{Tab:Wexpansion}
  \begin{tabular}{c c c}
    \hline
    $M_{AA}$ & $M_{BB}$ & $W$\\
    \hline
    $|\Phi^+_{AA}\rangle=\frac{|0\rangle\otimes|0\rangle + |1\rangle\otimes|1\rangle}{\sqrt{2}}$ & $|\Phi^+_{BB}\rangle=\frac{|0\rangle\otimes|0\rangle + |1\rangle\otimes|1\rangle}{\sqrt{2}}$ & $W^{}=\sum_{s,t}\beta^{++}_{s,t}{\tau}_s^T\otimes{\omega}_t^T$\\
    $|\Phi^-_{AA}\rangle=\frac{|0\rangle\otimes|0\rangle - |1\rangle\otimes|1\rangle}{\sqrt{2}}$ & $|\Phi^-_{BB}\rangle=\frac{|0\rangle\otimes|0\rangle - |1\rangle\otimes|1\rangle}{\sqrt{2}}$ & $W^{}=\sum_{s,t}\beta^{--}_{s,t}\tilde{\tau}_s^T\otimes\tilde{\omega}_t^T$\\
    $|\Phi^+_{AA}\rangle=\frac{|0\rangle\otimes|0\rangle + |1\rangle\otimes|1\rangle}{\sqrt{2}}$ & $|\Phi^-_{BB}\rangle=\frac{|0\rangle\otimes|0\rangle - |1\rangle\otimes|1\rangle}{\sqrt{2}}$ & $W^{}=\sum_{s,t}\beta^{+-}_{s,t}{\tau}_s^T\otimes\tilde{\omega}_t^T$\\
    $|\Phi^-_{AA}\rangle=\frac{|0\rangle\otimes|0\rangle - |1\rangle\otimes|1\rangle}{\sqrt{2}}$ & $|\Phi^+_{BB}\rangle=\frac{|0\rangle\otimes|0\rangle + |1\rangle\otimes|1\rangle}{\sqrt{2}}$ & $W^{}=\sum_{s,t}\beta^{-+}_{s,t}\tilde{\tau}_s^T\otimes{\omega}_t^T$\\
    \hline
  \end{tabular}
\end{table}

Next, we need to calculate the coefficients $\beta_{s,t}^{\pm\pm}$ and the corresponding probabilities $p(\pm,\pm|\tau_s,\omega_t)$ for given ancillary quantum states $\{\tau_s\}$ and $\{\omega_t\}$. Define $\sigma_0 = I$ and $\sigma_1, \sigma_2, \sigma_3$ to be the Pauli matrices. Then let $\tau_s$ and $\omega_s$ both be the eigenstates of $\sigma_s$ with eigenvalues of 1. That is, $\tau_0 = \omega_0 = I/2$, $\tau_s = \omega_s = (I + \sigma_s)/{2}$ for $s = 1, 2, 3$. By decomposing $W$ into $\{\tau_s^T\otimes\omega_t^T\}$ and $\{\widetilde{\tau}_s^T\otimes\widetilde{\omega}_t^T\}$, we find that the coefficients $\beta^{ab}_{st}$ and the probabilities $p(a,b|\tau_s,\omega_t)$ of the two cases $++$ and $--$ are the same, and those of $+-$ and $-+$ are the same.


In the cases of $++$ and $--$, the coefficients are given by
\begin{equation}
[\beta^{++}_{st}] = [\beta^{--}_{st}] = \left[
  \begin{array}{cccc}
    4 & -1 & -1 & -1 \\
    -1 & 1 & 0 & 0 \\
    -1 & 0 & 1 & 0 \\
    -1 & 0 & 0 & 1 \\
  \end{array}
\right],
\end{equation}
with corresponding probabilities of
\begin{equation}\label{}
  p(+,+|\tau_s,\omega_t) = p(-,-|\tau_s,\omega_t)=
  \left[
    \begin{array}{cccc}
      1/16 & 1/16 & 1/16 & 1/16 \\
      1/16 & (1-v)/16 & 1/16 & 1/16 \\
      1/16 & 1/16 & (1-v)/16 & 1/16 \\
      1/16 & 1/16 & 1/16 & (1-v)/8 \\
    \end{array}
  \right].
\end{equation}
There are ten nonzero terms in the coefficient matrix, so ten different ancillary inputs ($\tau_s, \omega_t$) are required. In practice, it is possible to reduce the number of inputs by introducing two other states $\tau_4 = \frac{I+(\sigma_x+\sigma_y+\sigma_z)/\sqrt{3}}{2}$ and $\omega_4 = \frac{I+(\sigma_x+\sigma_y+\sigma_z)/\sqrt{3}}{2}$. In this case, we have another decomposition of $W$ with coefficients of
\begin{equation}
[\beta^{++}_{st}] = [\beta^{--}_{st}] = \left[
  \begin{array}{ccccc}
    2\sqrt3-2 & 0 & 0 & 0 & -\sqrt3 \\
    0 & 1 & 0 & 0 & 0\\
    0 & 0 & 1 & 0 & 0 \\
    0 & 0 & 0 & 1  & 0\\
    -\sqrt3 & 0 & 0 & 0  & 0\\
  \end{array}
\right].
\end{equation}
In this setting, only six ancillary sets are required (comparing to ten in the original construction). As a result, we derive the coefficients and probabilities in Eq.~\eqref{Supp:MDIEWJ} for outcomes $++$ and $--$, as shown in Table \ref{Tab:Coeffppmm}.

\begin{table}[hbt]
\caption{Coefficients and probabilities for MDIEW with outcomes $++$ and $--$. Note that when $\beta=0$, the corresponding probability $p$ is irrelevant.} \label{Tab:Coeffppmm}
\begin{tabular}{c | c c c c c}
  \hline
   & $\tau_0= I/2$ & $\tau_1 = \frac{I+\sigma_x}{2}$ & $\tau_2 = \frac{I+\sigma_y}{2}$ & $\tau_3 = \frac{I+\sigma_z}{2}$ & $\tau_4 = \frac{I+(\sigma_x+\sigma_y+\sigma_z)/\sqrt{3}}{2}$  \\
   \hline
  $\omega_0 = I/2$ & $\beta = 2\sqrt3-2, p = \frac{1}{16}$ & $\beta = 0$ & $\beta = 0$ & $\beta = 0$ & $\beta = -\sqrt3, p = \frac{1}{16}$ \\
  $\omega_1 = \frac{I+\sigma_x}{2}$ & $\beta = 0$ & $\beta = 1, p = \frac{1-v}{16}$ & $\beta = 0$ & $\beta = 0$ & $\beta = 0$ \\
  $\omega_2 = \frac{I+\sigma_y}{2}$ & $\beta = 0$ & $\beta = 0$ & $\beta = 1, p = \frac{1-v}{16}$ & $\beta = 0$ & $\beta = 0$ \\
  $\omega_3 = \frac{I+\sigma_z}{2}$ & $\beta = 0$ & $\beta = 0$ & $\beta = 0$ & $\beta = 1, p = \frac{1-v}{8}$ & $\beta = 0$ \\
  $\omega_4 = \frac{I+(\sigma_x+\sigma_y+\sigma_z)/\sqrt{3}}{2}$ & $\beta = -\sqrt3, p = \frac{1}{16}$ & $\beta = 0$ & $\beta = 0$ & $\beta = 0$ & $\beta = 0$ \\
  \hline
\end{tabular}
\end{table}

Similarly, for the other two cases of outcomes $+-$ and $-+$, the coefficients are
\begin{equation}
[\beta^{-+}_{st}] = [\beta^{+-}_{st}] = \left[
  \begin{array}{cccc}
    0 & 1 & 1 & -1 \\
    1 & -1 & 0 & 0 \\
    1 & 0 & -1 & 0 \\
    -1 & 0 & 0 & 1 \\
  \end{array}
\right]
\end{equation}
with corresponding probabilities of
\begin{equation}\label{}
  p(+,-|\tau_s,\omega_t) = p(-,+|\tau_s,\omega_t)=
  \left[
    \begin{array}{cccc}
      1/16 & 1/16 & 1/16 & 1/16 \\
      1/16 & (1+v)/16 & 1/16 & 1/16 \\
      1/16 & 1/16 & (1+v)/16 & 1/16 \\
      1/16 & 1/16 & 1/16 & (1-v)/8 \\
    \end{array}
  \right].
\end{equation}
when using the ancillary states $\tau_0 = \omega_0 = I/2$, $\tau_s = \omega_s = (I + \sigma_s)/{2}$ for $s = 1, 2, 3$. Similarly, we can define $\tau'_4 = \frac{I+(-\sigma_x-\sigma_y+\sigma_z)/\sqrt{3}}{2},\, \omega'_4 = \frac{I+(-\sigma_x-\sigma_y+\sigma_z)/\sqrt{3}}{2}$ so that another decomposition of $W$ is derived,
\begin{equation}
[\beta^{+-}_{st}] = [\beta^{-+}_{st}] = \left[
  \begin{array}{ccccc}
    2\sqrt3+2 & 0 & 0 & 0 & -\sqrt3 \\
    0 & -1 & 0 & 0 & 0\\
    0 & 0 & -1 & 0 & 0 \\
    0 & 0 & 0 & 1  & 0\\
    -\sqrt3 & 0 & 0 & 0  & 0\\
  \end{array}
\right]
\end{equation}
Again, in this setting, only six measurements are required. The coefficients and probabilities of outcomes $+-$ and $-+$ are shown in Table \ref{Tab:Coeffpmpm}.

\begin{table}[hbt]
\caption{Coefficients and probabilities for MDIEW with outcomes $+-$ and $-+$. Note that when $\beta=0$, the corresponding probability $p$ is irrelevant.} \label{Tab:Coeffpmpm}
\begin{tabular}{c | c c c c c}
  \hline
   & $\tau_0= I/2$ & $\tau_1 = \frac{I+\sigma_x}{2}$ & $\tau_2 = \frac{I+\sigma_y}{2}$ & $\tau_3 = \frac{I+\sigma_z}{2}$ & $\tau'_4 = \frac{I+(-\sigma_x-\sigma_y+\sigma_z)/\sqrt{3}}{2}$  \\
   \hline
  $\omega_0 = I/2$ & $\beta = 2\sqrt3+2, p = \frac{1}{16}$ & $\beta = 0$ & $\beta = 0$ & $\beta = 0$ & $\beta = -\sqrt3, p = \frac{1}{16}$ \\
  $\omega_1 = \frac{I+\sigma_x}{2}$ & $\beta = 0$ & $\beta = -1, p = \frac{1+v}{16}$ & $\beta = 0$ & $\beta = 0$ & $\beta = 0$ \\
  $\omega_2 = \frac{I+\sigma_y}{2}$ & $\beta = 0$ & $\beta = 0$ & $\beta = -1, p = \frac{1+v}{16}$ & $\beta = 0$ & $\beta = 0$ \\
  $\omega_3 = \frac{I+\sigma_z}{2}$ & $\beta = 0$ & $\beta = 0$ & $\beta = 0$ & $\beta = 1, p = \frac{1-v}{8}$ & $\beta = 0$ \\
  $\omega'_4 = \frac{I+(-\sigma_x-\sigma_y+\sigma_z)/\sqrt{3}}{2}$ & $\beta = -\sqrt3, p = \frac{1}{16}$ & $\beta = 0$ & $\beta = 0$ & $\beta = 0$ & $\beta = 0$ \\
  \hline
\end{tabular}
\end{table}

Although each of the four cases above defines an MDIEW, we can combine four of them as one to enhance the successful probability of MDIEW,
\begin{equation} \label{eq:newJ}
\begin{aligned}
J &= \frac{1}{4}\sum_{a, b}\sum_{s,t}\beta^{a, b}_{s,t}p(a, b|\tau_s, \omega_t)\\
&= \frac{1}{4}\sum_{s,t}( \beta^{++}_{s,t}p(+, +|\tau_s, \omega_t) + \beta^{+-}_{s,t}p(+, -|\tau_s, \omega_t) + \beta^{-+}_{s,t}p(-, +|\tau_s, \omega_t) +\beta^{--}_{s,t}p(-, -|\tau_s, \omega_t))
\end{aligned}
\end{equation}
By doing this, we improve the efficiency of experiments by four times comparing to the original proposal \cite{Branciard13}.

To witness entanglement for the bipartite states defined in Eq.~\eqref{rhovab} with MDIEW defined in Eq.~\eqref{eq:newJ}, in total eight different ancillary state pairs should be prepared, and the results are summarized in Table \ref{decom}.
\begin{table}[hbt]
  \caption{Our MDIEW in the form of Eq.~\eqref{eq:newJ} for the bipartite states defined in Eq.~\eqref{rhovab}.}\label{decom}
  \begin{tabular}{cccccc}
    \hline
    $\tau_s$ & $\omega_t$ & $\beta^{++}_{st}=\beta^{--}_{st}$ & $p(+,+|\tau_s, \omega_t)=p(-,-|\tau_s, \omega_t)$ & $\beta^{+-}_{st}=\beta^{-+}_{st}$ & $p(+,-|\tau_s, \omega_t)=p(-,+|\tau_s, \omega_t)$\\
    \hline
    $I/2$ & $I/2$& $2\sqrt{3}-2$& $1/16$ & $2\sqrt{3}+2$& $1/16$\\
    $\frac{I+\sigma_x}{2}$ & $\frac{I+\sigma_x}{2}$& $1$& $(1-v)/16$& $-1$& $(1+v)/16$\\
    $\frac{I+\sigma_y}{2}$ & $\frac{I+\sigma_y}{2}$& $1$& $(1-v)/16$& $-1$& $(1+v)/16$\\
    $\frac{I+\sigma_z}{2}$ & $\frac{I+\sigma_z}{2}$& $1$& $(1-v)/8$& $1$& $(1-v)/8$\\
    $I/2$ & $\frac{I+(\sigma_x+\sigma_y+\sigma_z)/\sqrt{3}}{2}$& $-\sqrt{3}$& $1/16$ & 0 &- \\
    $\frac{I+(\sigma_x+\sigma_y+\sigma_z)/\sqrt{3}}{2}$ & $I/2$& $-\sqrt{3}$& $1/16$ &0&- \\
    $I/2$ & $\frac{I+(-\sigma_x-\sigma_y+\sigma_z)/\sqrt{3}}{2}$&0&- & $-\sqrt{3}$& $1/16$\\
    $\frac{I+(-\sigma_x-\sigma_y+\sigma_z)/\sqrt{3}}{2}$ & $I/2$&0&- & $-\sqrt{3}$& $1/16$\\
    \hline
  \end{tabular}
\end{table}

\section{Time-shift attack}
The idea of time-shift attack is originated from quantum cryptography \cite{qi07} and takes advantage of efficiency mismatches existing in measurement devices. Inspired by this idea, we construct a time-shift attack for the conventional witness defined in Eq.~\eqref{wit}.  Define $\sigma_0 = I$ and $\sigma_1, \sigma_2, \sigma_3$ be the Pauli matrices $\sigma_x$, $\sigma_y$, and $\sigma_z$, correspondingly. Then we can decompose $W$ to
\begin{equation}\label{}
  W = \frac{1}{4}(\sum_{i = 0}^3\sigma_i\otimes\sigma_i),
\end{equation}
and the EW can be realized by local measurements,
\begin{equation}\label{eq:expEW}
Tr[W\rho_{AB}]=\frac{1}{4}\left( {1 + \left\langle {{\sigma _x}{\sigma _x}} \right\rangle  + \left\langle {{\sigma _y}{\sigma _y}} \right\rangle  + \left\langle {{\sigma _z}{\sigma _z}} \right\rangle } \right).
\end{equation}

To realize the attack, we exploit the time mismatch of the two single-photon-counting modules (SPCMs) such that one detector is more efficient than the other. In this case, the real implementation ($W'$) is deviated from the original design witness $W$. In the attack Eve can suppress the positive contributes of the witness result $Tr[W\rho_{AB}]$ to let the witness result $Tr[W'\rho_{AB}]$ be negative by adjusting the time mismatch. For example, when measuring ${\sigma _x}{\sigma_x}$, Alice and Bob will project the input state to the eigenstates of $\sigma_x$, that is $\sigma_x^+$ and $\sigma_x^-$, corresponding to positive and negative eigenvalue respectively, and obtain probabilities $\left\langle {{\sigma _x^\pm}{\sigma _x^\pm}} \right\rangle$. Then the value of $\left\langle {{\sigma _x}{\sigma _x}} \right\rangle$ is defined as
\begin{equation}\label{}
  \left\langle {{\sigma _x}{\sigma _x}} \right\rangle = \left\langle {{\sigma _x^+}{\sigma _x^+}} \right\rangle + \left\langle {{\sigma _x^-}{\sigma _x^-}} \right\rangle - \left\langle {{\sigma _x^+}{\sigma _x^-}} \right\rangle - \left\langle {{\sigma _x^-}{\sigma _x^+}} \right\rangle.
\end{equation}
The probabilities $\left\langle {{\sigma _x^\pm}{\sigma _x^\pm}} \right\rangle$ is measured from coincidence counts $N_A^\pm N_B^\pm$ of detectors, that is
\begin{equation}\label{}
  \left\langle {{\sigma _x^\pm}{\sigma _x^\pm}} \right\rangle = \frac{N_A^\pm N_B^\pm}{\sum N_A^\pm N_B^\pm}.
\end{equation}
If the positive coincidence counts are all suppressed, that is $N_A^+ N_B^+ = N_A^- N_B^- = 0$, then the outcome of $\left\langle {{\sigma _x}{\sigma _x}} \right\rangle$ is
\begin{equation}\label{}
  \left\langle {{\sigma _x}{\sigma _x}} \right\rangle = - \left\langle {{\sigma _x^+}{\sigma _x^-}} \right\rangle - \left\langle {{\sigma _x^-}{\sigma _x^+}} \right\rangle = -\frac{N_A^+ N_B^-}{\sum N_A^\pm N_B^\pm} - \frac{N_A^- N_B^+}{\sum N_A^\pm N_B^\pm} = -1.
\end{equation}

Similarly, the all the other local measurements $\left\langle {{\sigma _y}{\sigma _y}} \right\rangle$ and  $\left\langle {{\sigma _z}{\sigma _z}} \right\rangle $ become $-1$ by suppressing positive coincidence counts, which gives a witness result of
\begin{equation}\label{}
  Tr[W'\rho_{AB}] = -\frac{1}{2}
\end{equation}
for any state $\rho_{AB}$.

In our experiment demonstration, we only suppress the positive coincidence counts to $10.9(1)\%$ instead of neglecting all of them to make a wrong witness result of a separable state to be entangled.

\section{Tomography}\label{appTomography}
\begin{figure} [hbt]
\centering
\resizebox{16cm}{!}{\includegraphics[scale=1]{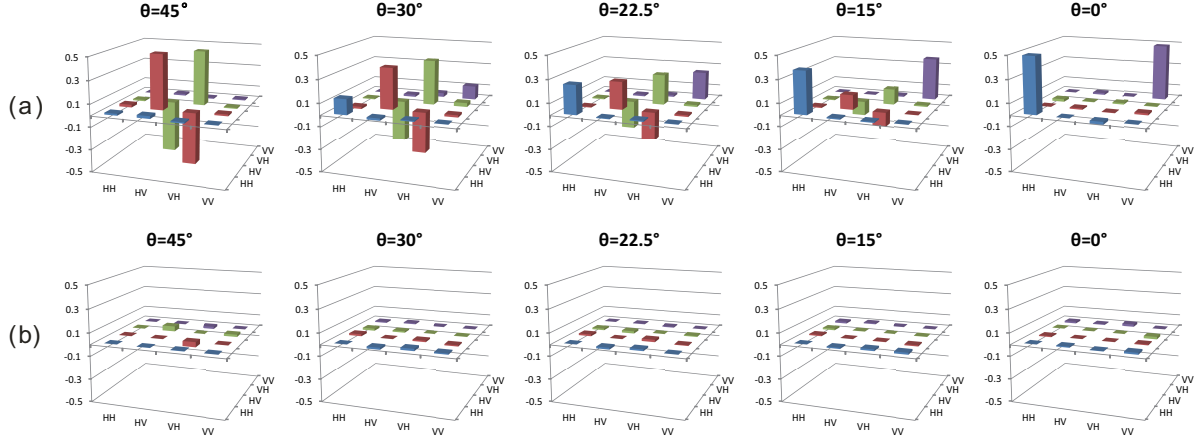}}
\caption{Tomography of the bipartite state $\rho^v_{34}$. Density matrices are constructed through tomography and over 250,000 coincidence detection events are obtained for each plot. Depending on the angle $\theta$ of the state selector defined in Eq.~(4) in Main Text, various states ${\rho _{34}^v}$ are prepared. (a) Real part of the density matrices $\rho^v_{34}$. (b) Imaginary part of the density matrices $\rho^v_{34}$.
}\label{denvt}
\end{figure}

In the experiment, we prepare the to-be-witnessed bipartite states $\rho_{AB}^v$ in the form of Eq.~\eqref{rhovab} with different values $v$. To verify whether the prepared states $\rho^v_{34}$ is close to the desired ones $\rho_{AB}^v$, their density matrices are reconstructed via quantum tomography with $v$ controlled by the angle $\theta$ of the selector HWP, as shown in Eq.~(4) in Main Text. The results of the density matrices are shown in Fig.~\ref{denvt}. Then we fit the value $v$ by the measured density matrixes $\rho^v_{34}$ to the desired states $\rho_{AB}^v$. As shown in Eq.~\eqref{rhovab}, $\rho_{AB}^v$ contains only real numbers, we can infer $v$ from the real part of $\rho^v_{34}$, and the imaginary parts are supposed to be near zero.

The parameter $v$ can be derived from the real-part of matrix $\rho^v_{34}$. For each matrix elements of $\rho^v_{34}$, $\rho_{11}, \rho_{22}, \rho_{33}, \rho_{44}$, and $\rho_{23}$ ($\rho_{32}$ is identical to $\rho_{23}$), one can estimate $v$, as shown in Table \ref{vfit1}. Accordingly, the average value of $v$ and its error bar are evaluated. As one can see that the experimental results agree the theoretical results well.

\begin{table}[hbt]
\caption{Tomography results of the input bipartite state.} \label{vfit1}
\begin{tabular}{cccccccccc}
    \hline
    & &\multicolumn{5}{c}{$v_{experiment}$} \\
    \cline{3-7}
     $\theta$ & $v_{theory}$ & $v_{\rho_{11}}$ & $v_{\rho_{22}}$ & $v_{\rho_{33}}$ & $v_{\rho_{44}}$ & $v_{\rho_{23}}$ & $\bar{v}_{exp}$ & $\delta\bar{v}_{exp}$ & $\delta v_{exp}$\\
    \hline
    45&    0&	0.0196 &	0.0228 &	0.0064 &	0.0258 	&0.0290 	&0.0207 	 &0.0039 	&0.0087\\
    30& 0.25&	0.2580 &	0.2538 	&0.2426 	&0.2686 	&0.2644 	&0.2575 	 &0.0045 	&0.0101\\
    22.5&  0.5	&0.4944 &	0.4820& 	0.4824 	&0.5230 	&0.5108 	&0.4985 	 &0.0081 	&0.0180\\
    15& 0.75&	0.7298 	&0.7198 	&0.7280 	&0.7718 	&0.7620 	&0.7423 	 &0.0103 	&0.0231\\
    0&    1&	0.9680 &	0.9818 &	0.9222 &	0.9684 	&0.9822 	&0.9645 	 &0.0110 	&0.0246\\
    \hline
  \end{tabular}
\end{table}


\section{Tangle}
To quantify the entanglement of quantum states, we adopt the measure of tangle \cite{Wootters}. For a 2-qubit state, $\rho_{AB}$, one can evaluate its tangle by the following steps.
\begin{enumerate}
\item
Define a non-Hermitian matrix
  \begin{equation}\label{}
    R = \rho_{AB}\Sigma\rho_{AB}^T\Sigma,
  \end{equation}
  where $\rho^T_{AB}$ is the transpose of $\rho_{AB}$, and the ``spin flip matrix $\Sigma$'' is defined as
  \begin{equation}\label{}
     \Sigma= \left[
  \begin{array}{ccccc}
0 &   0 &    0 &  -1\\		
0 &   0      &   1& 0	\\	
0  & 1  &   0        &  0	\\
-1&   0  & 0  & 0\\
  \end{array}
\right];
  \end{equation}

\item
Calculate the eigenvalues of $R$, and arrange them in decreasing order, $\lambda_1 \geq \lambda_2 \geq \lambda_3 \geq \lambda_4$;

\item
The concurrence of $\rho_{AB}$ is defined as
\begin{equation}\label{}
    C = \max\{0, \sqrt{\lambda_1}- \sqrt{\lambda_2}-\sqrt{\lambda_3}-\sqrt{\lambda_4}\};
\end{equation}

\item
The tangle is defined as
\begin{equation}\label{}
   tangle = C^2.
\end{equation}
\end{enumerate}

The tangle of a bipartite state is a measure of entanglement. If the tangle is zero, then the bipartite state $\rho_{AB}$ must be a separable state. For states defined in Eq.~\eqref{rhovab}, we can calculate the corresponding tangle. By following the aforementioned steps, we first calculate the four eigenvalues, $0, (1-v)^2, v^2/4, v^2/4$. For $v>2/3$, we have $v^2/4 > (1-v)^2$ and hence $tangle = C^2 = 0$. For $2/3\geq v$, we have $v^2/4 \leq (1-v)^2$ and hence $\sqrt{(1-v)^2}- 2\sqrt{v^2/4} = 1-2v$. Therefore, $C = 0$ for $v \geq 1/2$ and $C = 1-2v$ for $v<1/2$,
\begin{equation}\label{}
       tangle(\rho^v_{AB})= \left\{
  \begin{array}{cc}
(1-2v)^2 &  v<1/2 \\		
0 &   v\geq1/2. \\	
  \end{array}
  \right.
\end{equation}
The fitting value of $v$ from state tomography and the tangles are shown in Table \ref{ps}.


\begin{table}[hbt]
  \caption{The tangle values of the input states by tomography.}\label{ps}
  \begin{tabular}{cccccc}
  \hline
    $\theta_{exp}$& $v_{theory}$& $v_{exp}$ &  $v_{error}$& tangle($\rho^v_{34}(\theta)$) & $\textrm{tangle}_{\textrm{error}}$\\
    \hline
    $45^\circ$& 0 & 0.021 &	0.009 &	0.840 & 0.001\\
    $30^\circ$& 0.25 & 0.257 & 0.010 & 0.233 & 0.001\\
    $22.5^\circ$& 0.5 & 0.499 &	0.018 &	0 & 0\\
    $15^\circ$& 0.75 & 0.742 &	0.023 &	0 & 0\\
    $0^\circ$& 1 & 0.965 &	0.025 &	0 & 0\\
    \hline
  \end{tabular}\label{vtangle}
\end{table}

\end{appendix}


\bibliographystyle{apsrev4-1}

\bibliography{BibMDIEW}


\end{document}